\documentclass[review]{elsarticle}

\usepackage[latin1]{inputenc}

\usepackage{amsmath, amssymb}
\usepackage{natbib}
\usepackage{geometry}
\usepackage{fleqn}
\usepackage{txfonts}
\usepackage[hypertex]{hyperref}
\usepackage{scalefnt}
\usepackage{wrapfig}
\usepackage{graphicx}
\usepackage{pifont}
\usepackage{eurosym}
\usepackage{multirow}
\usepackage{booktabs}
\usepackage[dvips]{color}

\newtheorem{theorem}{Theorem}[section]
\newtheorem{lemma}[theorem]{Lemma}
\newtheorem{proposition}[theorem]{Proposition}

\newcommand{\nj}{N}
\newcommand{\qju}{Q_{1}}

\newcommand{\qjn}{Q_{\nj}}
\newcommand{\strike}{K}
\newcommand{\qj}{\qjn}
\newcommand{\qtju}{\widetilde{Q}_{1}}

\newcommand{\qtjn}{\widetilde{Q}_{\nj}}
\newcommand{\qwju}{Q^T_{1}}

\newcommand{\qwjn}{Q^T_{\nj}}
\newcommand{\qu}{q_1}

\newcommand{\qmu}{q_{-1}}

\newcommand{\cu}{c_1}

\newcommand{\cmu}{c_{-1}}

\newcommand{\pscad}{P}

\newenvironment{pf}{\ \\{\bf Proof:}}{\hfill\mbox{$\diamond$}\medskip}
\newenvironment{pfnome}{\ \\{\bf Proof}}{\hfill\mbox{$\diamond$}\medskip}

\journal{Applied Mathematics and Computation}

\begin{document}

\begin{frontmatter}

\title{Efficient European and American option pricing \\ under a jump-diffusion process}
\author{Marcellino Gaudenzi}
\author{Patrizia Stucchi}
\author{Alice Spangaro}
\address{Università di Udine, Dipartimento di Scienze Economiche e Statistiche,  via Tomadini 30/A, Udine}

\begin{abstract}
When the underlying asset displays oscillations, spikes or heavy-tailed distributions, the lognormal diffusion process (for which
Black and Scholes  developed their momentous option pricing formula) is inadequate: in order to overcome these real world difficulties many models have been developed. Merton proposed a jump-diffusion model, where the dynamics of the price of the underlying are subject to variations due to a Brownian process and also to possible jumps, driven by a compound Poisson process. Merton's model admits a series solution for the European option price, and there have been a lot of attempts to obtain a discretisation of the Merton model with tree methods in order to price American or more complex options, e. g. Amin, the $O(n^3)$ procedure by Hilliard and Schwartz and the $O(n^{2.5})$  procedure by Dai \textit{et al.}. Here, starting from the implementation of the seven-nodes procedure by Hilliard and Schwartz, we prove theoretically that it is possible to reduce the complexity to $O(n \ln n)$ in the European case and $O(n^2 \ln n)$ in the American put case. These theoretical results can be obtained through suitable truncation of the lattice structure and the proofs provide closed formulas for the truncation limitations.
\end{abstract}


\end{frontmatter}

\date{}
\thispagestyle{empty}

\section{Introduction}

As it is well known, Black and Scholes \cite{BS}  develop their momentous option pricing formula when the underlying asset is a stock and its price follows a lognormal diffusion process. However, frequently, financial derivatives cannot be priced by closed-form formulas and must be evaluated by numerical methods like trees or lattices (e.g. \cite{CRR}). Moreover, when the underlying asset displays oscillations, spikes or heavy-tailed distributions, the lognormal diffusion process is inadequate (see \cite{HS} for a detailed discussion).
In order to overcome these real world difficulties many models have been developed. One of the most popular models for its simplicity and efficiency has been proposed by Merton \cite{Merton76}. This is a jump-diffusion model, where the dynamics of the price of the underlying are subject to variations due to a Brownian process and also to possible jumps. Under the Black and Scholes \cite{BS} assumptions on the Brownian component, and considering a compound Poisson process for the jump part, the model admits a series solution for the European option price.

There have been several attempts to obtain a discretisation with tree methods of the Merton model with lognormal jumps in order to price American or more complex options.
Amin \cite{Amin}  proposes a procedure for derivative pricing by discretising the distribution of the underlying allowing the jumps to have a random amplitude which must be a multiple of the Brownian move. Hilliard and Schwartz \cite{HS}, considering the independency of the two processes involved in the Merton model, develop a multinomial lattice: one variable mimicking the diffusion process and the second one the lognormal jumps in the compound Poisson process. The HS procedure (when the second variable is allowed a seven-node branching at every time step) provides more accurate results than the one by Amin, and the weak convergence of the discrete price is ensured in the special case of deterministic jump amplitude and numerically justified otherwise. Hilliard and Schwartz apply their bivariate tree to the evaluation of American options, and  the time complexity of their backward procedure is $O(n^3)$. Dai \textit{et al.} \cite{Lyuu} build on the HS procedure reducing the complexity to $O(n^{2.5})$ by dissolving the intermediate nodes on the tree introduced by the jumps in the nearest diffusion nodes, therefore providing a one-dimensional tree. 

A common remedy among practitioners to overcome the computational cost of such a procedure is to neglect the tails of the distribution, computing the price only according to a part of the nodes of the tree, without taking into account those that are farthest from the expected value. What we propose here is to start from the HS technique and show that this practice can be theoretically founded: 
we prove that it is possible to reduce the complexity to $O(n \ln n)$ in the European case and $O(n^2 \ln n)$ in the American case through suitable truncation of the traditional lattice structure and the theoretical proofs provide closed formulas for the truncation limitations.
More in detail, 
we start from the evaluation of the European call option as the discounted expected payoff at maturity, derived from the HS backward procedure. Our basic idea consists in focusing on the jump probabilities relative to every ending node and analysing the error obtained considering only a range of cumulative jumps  proportional to $\ln n$.   We obtain an upper estimation of the error we make if we consider the backward procedure truncated throughout the whole tree with these limitations. This also provides an upper estimation of the error we get if we apply the truncation to the forward computation of the jump probabilities. In this way, we are able to construct a procedure of order $O(n \ln n)$ for the European case.  Finally, we move on to the American case proving that our tree gives in the put case an error less than the European one with respect to the HS procedure. The latter procedure is of order $O( n^2 \ln n)$.

The numerical results show that also for low values of the number of steps, our procedure improves significantly the existing techniques.
Our paper is organized as follows. Section 2 describes Merton's model. The Hilliard and Schwartz discretisation of the Merton model is introduced in Section 3.  Section 4 is devoted to the introduction of the main results of our truncation procedure, the European case is further developed in Section 5, while the American case is treated in Section 6.
Numerical results are given in Section 7. Section 8 contains the conclusions. With the aim of facilitating the reader, proofs of the (somewhat cumbersome) theoretical results for the procedure with a generic odd number of jump nodes  are included in the Appendix, available online.

\section{Merton model for jump-diffusion processes} 
\label{Merton}

The strong assumptions of the Black \& Scholes model do not always satisfy the real market dynamics; this reason has spurred the development of different models.

In particular, in order to describe the possibility of having, in the value of the underlying, significative variations in small amounts of time, which we will call \lq\lq jumps\rq\rq, Merton \cite{Merton76} provided a jump-diffusion model, where the dynamics of the price of the underlying are not only subjected to variations due to a Brownian process, but also to possible, if rare, greater variations, that can be caused by external events (e.g. arrival of information).

To model the random arrival of rare events, a Poisson distribution is used. Each event causes a random variation of the price,  determined by a random variable. The random variables that model the amplitude of the variation of the single events are supposed to be independent and identically distributed. Various distributions for the jump have been studied in literature, for reasons of simplicity and relevance  (see, for example, \cite{Kou}). We will consider a lognormally distributed amplitude of the jump, as in \cite{Merton76}, \cite{Amin}, \cite{HS}.

The underlying dynamics of the Merton model under these assumptions is given by the following equation:

	\begin{equation} \label{dyn} \frac{dS}{S} =(r-d - \lambda \bar{j}) dt + \sigma dz + Jdq \end{equation}

\noindent where $r$ is the risk-free interest rate, $d$ is the continuous dividend yield, $\sigma^2$ is the variance of the return provided that there are no jumps, $\lambda$ is the intensity of the Poisson process that models the arrival of jumps (i.e. $\lambda$ equals the average number of arrivals in a time unit), $dq$ assumes values 1 or 0 according to the presence or absence of a jump, $J$ is the random amplitude of the jump, such that $\ln(J+1) \sim N(\gamma', \delta^2)$, and $\bar{j}=E(J)$.

Applying Ito's Lemma in a formulation that also accounts for jumps (see \cite{itojump}) we can express the solution of Equation (\ref{dyn}) as:

	\begin{equation} \label{sol} S =S_0 e^{(r-d - \lambda \bar{j} - \frac{\sigma^2}{2})t + \sigma z(t)} \prod_{i=0}^{n(t)} (1+J_i) \end{equation}

\noindent where $n(t)$ is a Poisson process of parameter $\lambda$, $J_0=0$ and $\ln (1+J_i)  \sim N(\gamma', \delta^2) $ for $i \geq 1$.

In the following, for clarity purposes we will consider the logarithmic return as divided in two components $X_t$ and $Y_t$, where

\begin{equation*} X_t= \alpha t + \sigma z(t), \end{equation*}

with $\alpha = r-d - \lambda \bar{j} - \frac{\sigma^2}{2} $, is the diffusion component, while \begin{equation*}Y_t = \sum_{i=0}^{n(t)} \ln (1+J_i) \end{equation*} is the jump component, and we will focus on understanding the behaviour of the compound Poisson process $Y_t$.

\section{Hilliard and Schwartz's implementation} 
\label{HillSchw}

Hilliard and Schwartz \cite{HS} discretise Merton's model by creating a bivariate tree.
At every time step, the nodes of the tree model the values of the underlying return considering that this is influenced both by the Brownian motion and by the lognormal jumps.

Given $\tau$ as the time to maturity, $n$ the number of time steps, $\Delta t =\frac{\tau}{n}$ as the time interval, Hilliard and Schwartz consider $\sigma \sqrt{\Delta t}$ as the amplitude of the Brownian step, as per usual in binomial trees, and $h$ as the minimal amplitude of the jump. In order to recover a structure as faithful as possible to the jump dynamics, they introduce a node for the \lq\lq no jumps\rq\rq{} situation, and then additional nodes to take into consideration the possibility of a jump of amplitude $\pm h$, $\pm 2 h$, $\ldots$,  $\pm \nj h$. Hilliard and Schartz's choice for $h$ is $c\sqrt{\gamma'^2+\delta^2}$, with $0<c\leq 1$. A choice of $\nj=1$ gives a rough approximation of the underlying and therefore of the price of the derivatives; by choosing $\nj=2, 3, 4$ (that is, respectively, a five-, seven-, nine-node tree) the results are more refined. In the usual trade-off between precision and computational costs, the choice $\nj = 3$ (i.e. the seven-node tree) seems to be the best option \cite{HS}.

 Once  $\nj$ is fixed, at every interval there are two possibilities for the Brownian move ($+\sigma \sqrt{\Delta t}$ and $-\sigma \sqrt{\Delta t}$) and $2 \nj + 1$ possibilities for the jump. Thus, from every node depart  $2(2 \nj + 1)$ branches: one for every possible combination of the diffusion move and the jump move.

Every node of the tree can be labelled with a triplet $(i,j,l)$, where the first index $i$ keeps track of the time ($0 \leq i \leq n$), $j$ describes the effect of the Brownian moves up to that time ($0 \leq j \leq i$) and $l$ the result of the jump moves ($-\nj i \leq l \leq \nj i$).

$X_n$ and $Y_n$ (the discrete counterparts to the continuous random variables $X_\tau$ and $Y_\tau$) are the algebraic sum of $n$ i.i.d. processes $X_{\Delta}$ and $Y_{\Delta}$ respectively, where
$X_{\Delta}$  is the classical CRR discretisation of the Brownian motion, with the Nelson-Ramaswamy modification of the probabilities:

\begin{equation*}
X_{\Delta}:=\begin{cases}
+\sigma \sqrt{\Delta t} & \text{ with probability $p=\frac{1}{2} \left(1+  \frac{ \alpha \sqrt{\Delta t}}{\sigma} \right)  $}\\
-\sigma \sqrt{\Delta t}  & \text{with probability $1-p$}
\end{cases}
\end{equation*}

while $Y_{\Delta}$  is a discretisation of the amplitude of the jump:

\begin{equation*}
Y_{\Delta}:=\begin{cases}
lh & \text{ with probability $\lambda \Delta t \cdot q_l $ for $-\nj \leq l \leq  \nj$,}\\
0  & \text{with probability $1- \lambda \Delta t$.}
\end{cases}
\end{equation*}

The probabilities $q_l $ for $-\nj \leq l \leq  \nj$ are chosen such that the first $2\nj-1$ moments of $Y_{\Delta}$ match those of  $Y_{\Delta t}$, approximated by its cumulants.

For $p$ to be well defined, we need to impose $-1 \leq \frac{ \alpha \sqrt{\Delta t}}{\sigma} \leq 1$.

Let us denote by $S(i,j,l)$ the value of the underlying on the node  $(i,j,l)$: $S(i,j,l)= S_0 e^{(-i+2j) \sigma \sqrt{\Delta t} + lh}$.

The price of the option in the European case can be evaluated as the discounted expected value on all the possible payoffs at maturity, but it is more profitable to us to focus on the effects of the Brownian motion and the compound Poisson jumps separately.

Let us consider only the Poisson moves and name $\qj(l)$ the probability of a net balance of $l$ cumulative jumps at maturity, i. e. the probability for the discrete process $Y_n$ of taking value $-\nj n \leq l \leq \nj n$.
This is equal to stating that $\qj(l)$ is the probability of reaching any of the terminal nodes $(n, j, l)$ for $ 0 \leq j \leq n$. 

Though we do not have at our disposal a formula for $\qj$, we can compute it recursively, with a $O(n^2)$ procedure.

Similarly, let $\pscad(j)$ be the probability of a net balance of $j$ Brownian moves at maturity, i.e.
\begin{equation*} \pscad(j) = \binom{n}{j} p^j (1-p)^{n-j}.\end{equation*}

The European call option price  with strike $\strike$ is then   

\begin{equation} \label{euroV} V = e^{-r\tau} \sum_{l=-n \nj }^{n  \nj} \sum_{j=0}^{n} (S_0 e^{(-n+2j)\sigma \sqrt{\Delta t}+hl} - \strike)^+ \pscad(j)  \qj (l). \end{equation}

The backward procedure for the European and for the American call option price respectively are obtained via the recursion formula

\begin{equation} \label{recEur}
 V_E(i,j,l)=e^{-r\Delta t} \sum_{k = -\nj}^{\nj} (V_E(i+1, j+1, l+k) p + V_E(i+1, j, l+k) (1-p)) q_k
\end{equation}

\begin{equation} \label{recAme}
 V_A(i,j,l)= \max \left \{ e^{-r\Delta t} \sum_{k = -\nj}^{\nj} (V_A(i+1, j+1, l+k) p + V_A(i+1, j, l+k) (1-p)) q_k, S(i,j,l) -\strike \right \}
\end{equation}

with initial data $V_E(n,j,l)=V_A(n,j,l)=(S(n,j,l)-\strike)^+$,  for $j$ integer between $0$ and $n$ and $l$ integer between $- \nj n$ and $\nj n$.

We will use the same notations   for the European put option price computed as the discounted expected value at maturity and the backward procedure for the European and the American put option prices respectively, with initial data $V_E(n,j,l)=V_A(n,j,l)=(\strike - S(n,j,l))^+$,  for $j$ integer between $0$ and $n$ and $l$ integer between $- \nj n$ and $\nj n$.

In the European case, the pricing obtained via the discounted expected value  and that obtained via the backward procedure coincide, that is $V=V_E(0,0,0)$.

\medskip

\section{Main results}

\subsection{Cutting the tree}  
\label{cutting}
																										
In this section we propose a 
$O(n \ln n)$ and a $O(n^2 \ln n)$ procedure respectively for the evaluation of European and American option prices.

Given two positive integers $\overline{k}$,$\overline{l} \leq \nj n $  we will call $\qwjn(l)$, with $-\nj n \leq l \leq  \nj n$, the probability (computed recursively forward) of reaching level $l$ of cumulative jumps at maturity without going out of the borders $-\overline{l}$ and $\overline{k}$ at any time step. 

We will focus on the value $V^{TT}$ defined as the discounted average of the option values restricted to the terminal nodes $(n,j,l)$ with $-\overline{l} \leq l \leq \overline{k}$, each weighed with probability $\pscad(j)\qwjn(l)$. For the call option, this means 

\begin{equation}
\label{vethat}
 	V^{TT} = e^{-r\tau} \sum_{l=-\overline{l}}^{\overline{k}} \sum_{j=0}^{n} (S_0 e^{(-n+2j)\sigma \sqrt{\Delta t}+hl} - \strike)^+ \pscad(j) \qwjn(l).
	\end{equation}

If we truncate the tree as described above, we are losing probability contributions in two different ways:
\begin{description}
\item [(a)] neglecting the paths that would reach - at maturity - a node outside the allowed region, i.e. a  node $(n, j, k)$ with  $ k > \overline{k}$ or  $ k < -\overline{l}$, for any $0 \leq j \leq n$;
\item [(b)] neglecting the paths that, even though ending at maturity in a node inside the allowed region, have at some point before maturity trespassed at least one of the boundaries.
\end{description}

Let us indicate with $V^T$ the value obtained by only using type $\textbf{(a)}$ elimination. For the call option this will be:

\begin{equation}
\label{HSN}
 V^T = e^{-r\tau} \sum_{k=-\overline{l}}^{\overline{k}} \sum_{j=0}^{n} (S_0 e^{(-n+2j)\sigma \sqrt{\Delta t}+hk} - \strike)^+ \pscad(j)  \qjn (k). \end{equation}

The following is true:

\begin{equation*}V^{TT} \leq V^T \leq V.\end{equation*}

We will indicate with $\qtjn (l) $ the value that differs from probability $\qjn (l)$ in that the single jump probabilities $q_{+i}$ and $q_{-i}$ are both substituted by the maximum of the two, for $ 1 \leq i \leq \nj$.
Obviously, $\qtjn $ can be computed recursively forward with an $O(n^2)$ procedure and $\qjn (\pm l)  \leq \qtjn (l)= \qtjn (-l) $.

We will show that we can choose $\overline{l}$ and $\overline{k}$  such that the error $ V - V^{TT} $ is less than an arbitrary $\varepsilon$, and show that for $\varepsilon = \frac{1}{n}$ we obtain  $\overline{l}$ and $\overline{k}$ that are proportional to $\ln n$. In this case $\qwjn$ can be computed recursively with an $O(n \ln n)$ procedure, therefore we can price the European option value via $V^{TT}$ with an  $O( n \ln n ) $ procedure. For the American option case we will define a backward procedure that is also confined between levels $-\overline{l}$ and $\overline{k}$, therefore an $O(n^2 \ln n)$ procedure. 
Our main purpose is to explicit analytical estimates of $\overline{l}$ and $\overline{k}$.

\subsection{Results}

Since the probability $q_i$ of a jump of amplitude $ih$ for $i$ integer, $-\nj \leq i \leq \nj$, in a $\Delta t $ time interval, is determined imposing a moment-matching condition, with the moments approximated by the cumulants, which are time-homogeneous, all $q_i$'s with $i \neq 0$ are inversely proportional to $n$. We will denote:

\begin{description}
\item[$c_i$] as the constant such that $q_i= \frac{c_i}{n}$ for $i \not= 0$, $-\nj \leq i \leq \nj$;
\item[$w_i$] as the maximum between $c_{+i}$ and $c_{-i}$ for $1 \leq i \leq \nj$;
\end{description}
and define recursively
\begin{gather} \label{iVudoppi}
\begin{split}
 W_1 & = w_1 \\
W_{i+1} & = w_{i+1} + W_i^{\frac{i+1}{i}}.
\end{split}
\end{gather}
 and \begin{equation*}M_i=\max \left \{W_i, W_i^{\frac{-i+1}{i}} \right \}.\end{equation*}

Let us denote
$G =  2 \nj \max \{ W_{\nj}, 1 \}  e^{W_{\nj}} \prod_{i= 1}^{\nj -1} M^2_i  $.

\begin{theorem} \label{errorbacknj}
Given $\varepsilon >0 $, considering $V$ the HS European call option value, taking 
\begin{align}
  \overline{k} & \geq \max \{ \nj \left \lceil e^{h\nj +1 }W_{\nj} - \ln \varepsilon + \ln( 4 S_0 G ) + (\alpha-r) \tau + \ln k_+ \right \rceil -1 ,  \nj \left \lceil 2 e^{h\nj}W_{\nj} - 1 \right \rceil -1 \} \\
 \overline{l} & \geq \max \{ \nj \left \lceil e^{-h\nj +1 }W_{\nj} - \ln \varepsilon + \ln( 4 S_0 G ) + (\alpha-r) \tau + \ln k_- \right \rceil -1 ,  \nj \left \lceil 2e^{h\nj}W_{\nj} - 1 \right \rceil -1 \}
\end{align}
with
\begin{align*}
k_+& = \sum_{r=0}^{\nj-1}  e^{hr}    +    \nj \max \{ W_{\nj}^2, 1 \}  e^{2h \nj } \sum_{r=0}^{\nj-1}  e^{-hr} \\
k_- & = \sum_{r=0}^{\nj-1}  e^{-hr} +  \nj \max \{ W_{\nj}^2, 1 \} \sum_{r=0}^{\nj-1}  e^{hr},
\end{align*}
 we have that the European call option value  $V^{TT}$ obtained via truncation of the tree at levels  $\overline{k}$ and $-\overline{l}$ 
satisfies:
\begin{equation*} V- V^{TT}  < \varepsilon\end{equation*}
\end{theorem}

\begin{theorem} \label{putnj}
Given $\varepsilon >0 $, considering $V$ the HS European put option value,  taking $\overline{k} \geq \max \{  \nj  \lceil 2 W_{\nj} - 1 \rceil -1 ,  \nj  \lceil W_{\nj} e -\ln \varepsilon - r \tau + \ln(4 \nj  (\nj+1) K G)  \rceil -1 \}$, we have that the  European put option value  $V^{TT} $ obtained via truncation of the tree at levels  $\overline{k}$ and $-\overline{l}$ with $\overline{l}=\overline{k}$ satisfies
\begin{equation*}  V- V^{TT} < \varepsilon\end{equation*}
\end{theorem}

\begin{theorem} \label{complexEuronj}
Given $\varepsilon=\frac{1}{n} >0 $, and $\overline{k}, \overline{l}$ the smallest integers  as in Theorem \ref{errorbacknj} in the call case and as in Theorem \ref{putnj} for the put case, the proposed procedure for $V^{TT}$ converges to the HS price and its computational complexity is  $O(n \ln n)$. 
\end{theorem}

Given a fixed $b>0$, we define the value $V_E^b(0,0,0)$ obtained via backward procedure according to the following formula:

\begin{equation*}\label{web}
V_E^b(i,j,k)=\begin{cases}
e^{-r\Delta t} \sum_{l = -\nj}^{\nj} (V_E^b(i+1, j+1, k+l) p + V_E^b(i+1, j, k+l) (1-p)) q_l  & \text{if $k \in [-\overline{l}, \overline{k}]$,}\\
b & \text{otherwise,}
\end{cases}
\end{equation*}
 with initial data in the call case 
\begin{equation*}
V_E^b(n,j,k)=\begin{cases}
(S(n,j,k)-\strike)^+  & \text{if $k \in [-\overline{l}, \overline{k}]$,}\\
b & \text{otherwise,}
\end{cases}
\end{equation*}
and initial data in the put case
\begin{equation*}
V_E^b(n,j,k)=\begin{cases}
(\strike-S(n,j,k))^+  & \text{if $k \in [-\overline{l}, \overline{k}]$,}\\
b & \text{otherwise.}
\end{cases}
\end{equation*}

The American value $V_A^b(0,0,0)$  obtained via backward procedure on a tree where the value of the option in any node outside the allowed region is substituted by $b$ is given by the following recursion pattern: 
$C_A^b(i,j,k) =e^{-r\Delta t} \sum_{l = -\nj}^{\nj} (V_A^b(i, j+1, k+l) p + V_A^b(i, j-1, k+l) (1-p)) q_l$  for all $i=0, \ldots, n$, $j=0, \ldots, i$, $k = -\nj i, \ldots, \nj i$ while 
$ V_A^b(i,j,k) = \max ( C_A^b(i,j,k) , S(i,j,k)-\strike)$ for the call option and $ V_A^b(i,j,k) = \max ( C_A^b(i,j,k) , \strike-S(i,j,k))$ for the put option  if $k \in [-\overline{l}, \overline{k}]$, $b$ otherwise; with initial data $V_A^b(n,j,k)=V_E^b(n,j,k)$. 

Taking $b=\strike$, we can state the following results.

\begin{theorem} \label{American}
Let $V_A(0,0,0)$ and $V_E(0,0,0)$ the binomial prices, evaluated with the backward procedure of Hilliard and Schwartz, in the  American put case and the European put case. Fixed $\overline{k}$ and $\overline{l}$, let $V_A^K(0,0,0)$ and $V_E^K(0,0,0)$  the binomial prices, evaluated with the backward procedure with the truncation described above, respectively in the American put case and in the European put case. One has:
$$|V_A^K(0,0,0)-V_A(0,0,0)|\leq |V_E^K(0,0,0)-V_E(0,0,0)|.$$
\end{theorem}

 \begin{theorem} \label{ameback}
Given $\varepsilon=\frac{1}{n} >0 $, and $\overline{k}=\overline{l}$ the smallest integer such that $\overline{k} \geq \max \{  \nj  \lceil 2 W_{\nj} - 1 \rceil -1 ,  \nj  \lceil W_{\nj} e -\ln \varepsilon  + \ln(4 \nj  (\nj+1) \strike G)  \rceil -1 \}$,
the backward procedure described above for the put price $V_A^K(0,0,0)$ converges to the HS price and its computational complexity is  $O(n^2 \ln n)$.
\end{theorem}

\section{European option} 

Our procedure will be detailed firstly for the European option. The American put option case is treated afterwards, and the reasoning relies on the results of the European case. 
For the sake of simplicity, we illustrate our procedure in the case $\nj = 1$, which can clarify the strategy adopted. The proofs in the case of an arbitrary $\nj$ are contained in the Appendix.

\subsection{Some preliminary results}

These inequalities will be used in the following.

\begin{proposition} \label{brow}
\begin{equation*} e^{-r\tau}  \sum_{j=0}^{n}  e^{\sigma\sqrt{\Delta t }(-n+2j)}   \pscad(j)  \leq  e^{(\alpha-r)\tau} \end{equation*}
\end{proposition}

\begin{pf}

We can write

\begin{align}
\sum_{j=0}^{n} \binom{n}{j} e^{\sigma\sqrt{\Delta t} (-n+2j)}  p^j (1-p)^{n-j}  & =  \sum_{j=0}^{n} \binom{n}{j}  (e^{\sigma\sqrt{\Delta t} } p)^j [ e^{-\sigma\sqrt{\Delta t} }  (1-p)]^{n-j}  \\
& = [ e^{\sigma\sqrt{\Delta t}} p +  e^{-\sigma\sqrt{\Delta t} }(1-p) ]^n
\end{align}

Since the value $e^{\sigma\sqrt{\Delta t }}p +  e^{-\sigma\sqrt{\Delta t }}(1-p)$ is higher the higher the probability $p$, the worst case scenario (since we would like to find an upper bound) is $p=1$.

Since $p$ is defined as $\frac{1}{2} \left(1+  \frac{ \alpha \sqrt{\Delta t}}{\sigma} \right)  $, this means $\alpha \Delta t = \sigma \sqrt{\Delta t}$, therefore:

\begin{align*}
 \sum_{j=0}^{n} \binom{n}{j} e^{\sigma\sqrt{\Delta t }(-n+2j)}  p^j (1-p)^{n-j}  & \leq (e^{\sigma\sqrt{\Delta t} } )^n  \\
& \leq ( e^{\alpha \Delta t } )^n = e^{\alpha \tau }
\end{align*}

\end{pf}

\begin{lemma} \label{resto}

If $ \displaystyle 0 \leq x \leq \frac{n+1}{j}$ for some $j,n \in \mathbb{N}$, $j >1$,  then $ \displaystyle  \sum_{i \geq n} \frac{x^i}{i!}  \leq \frac{j}{j-1}  \frac{x^n}{n!}$.
\end{lemma}

\begin{pf}
\noindent  We consider $ \displaystyle  \sum_{n=0}^{\infty} a_n$.  If the terms of the summation are such that $ \displaystyle  a_{i+1} \leq \frac{1}{j} a_i$ $\forall i$ and $a_i \geq 0$, then \begin{equation*} \displaystyle  \sum_{i=0}^{\infty} a_i \leq a_0 \sum_{i=0}^{\infty} \frac{1}{j^i} = \frac{j}{j-1} a_0.\end{equation*}

\noindent We apply this to the situation $a_i = \frac{x^i}{i!}$.

\noindent If $ \displaystyle  0 \leq x \leq \frac{n+1}{j}$  then we have $ \displaystyle  a_{i+1} = \frac{x^{i+1}}{(i+1)!}=\frac{x^i}{i!} \frac{x}{i+1} = a_i \frac{x}{i+1} \leq a_i \frac{n+1}{(i+1)j}$, which gives

\begin{equation*} a_{i+1} \leq   \frac{1}{j} a_i \end{equation*}

for $i \geq n$.

\noindent Therefore, \begin{equation*}\sum_{i \geq n} \frac{x^i}{i!} = \sum_{i=0}^{+\infty} \frac{x^{n+i}}{(n+i)!} \leq \frac{j}{j-1} \frac{x^n}{n!} \end{equation*}

\end{pf}

\begin{lemma} \label{logari}

Given $c> 0$ and $n\in \mathbb{N}$, $n \neq 0$,

\begin{equation*}\ln \frac{c^n}{n!} < n (\ln c+1) - n\ln n < -n + ce \end{equation*}
\end{lemma}

\begin{pf}

We can write $\ln \frac{c^n}{n!}$ as $n \ln c - \ln n!$.

By the Stirling series of $ \displaystyle  \ln n! = n \ln n -n +\frac{\ln (2\pi n)}{2} +\frac{1}{12n} - \frac{1}{360n^3} + \ldots $, and remembering that the error committed by truncating the series is of the same sign of the first term omitted, we have:

\begin{equation*}\ln n! > n \ln n -n +\frac{\ln (2\pi n)}{2} > n \ln n -n \end{equation*}

\noindent Therefore

\begin{equation*}n \ln c - \ln n!  < n \ln c - n \ln n+ n =n (\ln c+1) - n\ln n \end{equation*}

We set $a= \ln c +1$ and consider the function $f(x)=xa - x \ln x$. This is a concave function, therefore its graph lies below the tangent line in $x=e^{a}$. Since the derivative of $f$ is $f'(x)=a-1-\ln(x)$, $f(e^{a})=ae^{a}-a e^{a}=0$ and $f'(e^{a})=a-1-a=-1$, the equation of the tangent line is $y=-x+e^a$ and we get the following inequality:

\begin{equation*}na - n \ln n \leq -n+e^{a} = -n + e^{\ln c +1} = -n + ce \end{equation*}

hence the thesis.

\end{pf}

We will show that we can choose $\overline{k}$ and $\overline{l}$ such that computing 	$V^{TT} $ is less expensive than $V$ and the error $V - 	V^{TT} $ is less than an arbitrary $\varepsilon$.

Our strategy consists in splitting this difference in two components, $V -V^T$ and $V^T - V^{TT}$, and for each difference finding an upper estimate in terms of $\overline{l}$ and $\overline{k}$.

First of all, in treating both components we need to establish an upper estimate for the contribution of the Brownian motion to the possible values of the underlying.

Applying Proposition \ref{brow} with $\overline{k}, \overline{l} > 0$, we have the following limitation for the difference $V - V^T$ in the call case:

\begin{align}
  \notag V - V^T  = & e^{-r\tau} \left[\sum_{k=- \nj n }^{-\overline{l}-1} \sum_{j=0}^{n} (S_0 e^{(-n+2j)\sigma \sqrt{\Delta t}+hk} - \strike)^+ \pscad(j)  \qjn (k)  + \sum_{k=\overline{k}+1}^{\nj n} \sum_{j=0}^{n} (S_0 e^{(-n+2j)\sigma \sqrt{\Delta t}+hk} - \strike)^+ \pscad(j)  \qjn (k) \right]
\\
\notag  & \leq e^{-r\tau}\left[ \sum_{k=\overline{k}+1}^{ \nj n}  S_0  e^{hk} \sum_{j=0}^{n} e^{\sigma\sqrt{\Delta t }(-n+2j)}  \pscad(j) \qjn(k)  +  \sum_{k=\overline{l}+1}^{\nj n}  S_0  e^{-hk} \sum_{j=0}^{n} e^{\sigma\sqrt{\Delta t }(-n+2j)}  \pscad(j) \qjn(-k)\right]  \\
\notag  &  \leq e^{-r\tau}  S_0   \sum_{j=0}^{n} e^{\sigma\sqrt{\Delta t }(-n+2j)}  \pscad(j) 	\left( \sum_{k=\overline{k}+1}^{ \nj n}  e^{hk} \qtjn(k)  + \sum_{k=\overline{l}+1}^{ \nj n}  e^{-hk}  \qtjn(k) \right)  \\
 \label{tagliofuori} &  \leq e^{(\alpha-r)\tau}  S_0 \left( \sum_{k=\overline{k}+1}^{ \nj n}  e^{hk} \qtjn(k)  + \sum_{k=\overline{l}+1}^{\nj n}  e^{-hk}  \qtjn(k) \right).
\end{align}

\noindent In the put case, since the value of the option is smaller than the strike, we have:
\begin{equation}
    \label{tagliofuoriput} V - V^T   
  \leq e^{-r\tau} \strike   \sum_{j=0}^{n}  \pscad(j) 	\left( \sum_{k=\overline{k}+1}^{ \nj n} \qtjn(k)  + \sum_{k=\overline{l}+1}^{ \nj n} \qtjn(k) \right)   \leq e^{-r\tau} \strike \left( \sum_{k=\overline{k}+1}^{ \nj n}  \qtjn(k)  + \sum_{k=\overline{l}+1}^{\nj n} \qtjn(k) \right)
\end{equation}

In order to show that the difference between $V^{TT}$ and $ V^T$ is arbitrarily small, we need to understand the difference between $ \qwjn (k)$ and $\qjn (k)$ for $-\overline{l} \leq k \leq  \overline{k}$, which is made of the probabilities of the paths that reach $k$ at maturity having previously gone outside the $[ - \overline{l}, \overline{k} ]$ region.

For the sake of simplicity, we consider separately the probability of reaching $k$ having gone over the $\overline{k}$ level and the probability of reaching $k$ having gone under the $-\overline{l}$ level; the sum of the two is obviously greater than the quantity we want to estimate.

Fixed $\overline{k}$ and  $\overline{l}$, let $- \overline{l} \leq k \leq \overline{k}$. Let us call $\qjn^{\overline{k}}(k)$ the probability of a net balance of $k$ jumps at maturity while reaching at some point a net balance higher than $\overline{k}$  and ${\qjn}_{\overline{l}}(k)$ the probability of a net balance of $k$ jumps at maturity while reaching at some point a net balance lower than $-\overline{l}$.

According to these definitions and by Proposition \ref{brow}, the difference between the values $V^T$ and $V^{TT}$ in the call case becomes:

\begin{align}
 V^T-	V^{TT}   & = e^{-r\tau} \sum_{k=-\overline{l}}^{\overline{k}} \sum_{j=0}^{n} (S_0 e^{(-n+2j)\sigma \sqrt{\Delta t}+hk} - \strike)^+ \pscad(j) (  \qjn (k)- \qwjn(k) ) \\
 & \leq e^{-r\tau}  S_0  \sum_{j=0}^{n} e^{\sigma\sqrt{\Delta t }(-n+2j)}  \pscad(j) \sum_{k=-\overline{l} }^{\overline{k}} e^{hk}  (  \qjn (k)- \qwjn(k) )\\
 \label{tagliodentro} & \leq   e^{(\alpha-r)\tau}  S_0  \sum_{k=-\overline{l} }^{\overline{k}} e^{hk} (\qjn^{\overline{k}} (k) +  {\qjn}_{\overline{l}} (k)  ).
\end{align}

In the put case, as in Equation \ref{tagliofuoriput} we have:
\begin{equation}  \label{tagliodentroput}
 V^T-	V^{TT} 
 \leq e^{-r\tau}  \strike \sum_{k=-\overline{l} }^{\overline{k}} (  \qjn (k)- \qwjn(k) ) 
 \leq   e^{-r\tau}  \strike  \sum_{k=-\overline{l} }^{\overline{k}} (\qjn^{\overline{k}} (k) +  {\qjn}_{\overline{l}} (k)  ).
\end{equation}

\subsection{European option, $\nj=1$} \label{eurocase1}

When $\nj=1$, the only possible values for the jump in a $\Delta t$ period are $-h$, $0$, $h$, with probabilities $\qmu, q_0, \qu$, which are the solutions of the following linear system:

\begin{equation*}
\begin{pmatrix}
1 & 1 & 1 \\
-1 & 0 & 1 \\
 1 & 0 & 1
\end{pmatrix}
\begin{pmatrix}
\qmu \\
q_0 \\
\qu
\end{pmatrix}
=
\begin{pmatrix}
1 \\
\frac{k_1}{h} \\
\frac{k_2}{h^2}
\end{pmatrix}
\end{equation*}

\noindent where $k_1=\lambda \Delta t \gamma'$ and $k_2 = \lambda \Delta t (\gamma'^2+\delta^2)=\lambda \Delta t h^2$ are the first two cumulants of the compound Poisson distribution (the usage of cumulants instead of moments has already been justified in \cite{HS}).

We can write:
\begin{align*} 
\qmu & =\frac{\lambda \tau }{2n}\left(1-\frac{\gamma'}{h} \right)=\frac{\cmu}{n} \\
q_0 & =1-\frac{\lambda \tau}{n} \\
\qu & =\frac{\lambda \tau}{2n}\left(1+\frac{\gamma'}{h} \right)=\frac{\cu}{n}
\end{align*}

Given $n$ the number of steps, we consider the probability $\qju(k)$ 
 of a net balance of $k \geq 0$ jumps at maturity:

\begin{equation*} \qju (k) = \sum_{t=0}^{\left \lfloor \frac{n-k}{2} \right \rfloor} \frac{n!}{(n-k-2t)!(k+t)!t!} \qu^{k+t} \qmu^t q_0^{n-k-2t} \end{equation*}

which is the sum of all the probabilities of the \lq\lq jump paths\rq\rq{} with $k+t$ up jumps and $t$ down jumps, for $t$ such that $k+2t \leq n$.

Since $\qu = \frac{ \cu}{n}$,  $\qmu = \frac{ \cmu}{n}$, let us define $w=\max \{ \cu, \cmu \}$;  the \lq\lq enlarged probability\rq\rq{}  $\qtju (k)$  is given by:

\begin{equation} \label{qtju}
\qtju (k) = \sum_{t=0}^{\left \lfloor \frac{n-k}{2} \right \rfloor} \frac{n!}{(n-k-2t)!(k+t)!t!}  \left( \frac{w}{n} \right)^{k+2t} q_0^{n-k-2t}.
\end{equation}

We are going to prove that $\qtju (k)$  is negligible for $ \lvert k \rvert > a \ln n +b $ (for some fixed constants $a$ and $b$), and that this translates into the negligibility of some of the possible payoffs in the evaluation of the option.

Consider Eq. (\ref{tagliodentro}) and (\ref{tagliodentroput}) with $\nj=1$; in order to deal with them, we relate  $\qju^{\overline{k}} (k)$ and ${\qju}_{\overline{l}} (k)$ to our \lq\lq enlarged probability\rq\rq{}  $\qtju(t)$.

\begin{lemma} \label{rientro1}
\begin{equation*}\qju^{\overline{k}}(k) \leq \qtju(2 \overline{k} -k+2)\end{equation*}
\begin{equation*}{\qju}_{\overline{l}}(k) \leq   \qtju(2 \overline{l} +k+2)\end{equation*}
for all $-\overline{l} \leq k \leq \overline{k}.$
\end{lemma}

\begin{pf}
We remark that the number of paths that reach the  $\overline{k}+1$ level at some point before maturity and end at  level $k$ is the same as the number of paths that end at level $2 \overline{k} -k+2$, by reflection principle (see \cite{Fel}).

Our intent is to recover an upper estimate of $\qju^{\overline{k}}(k)$ using the probability of the \lq\lq reflected\rq\rq{} paths.

We consider a single path that reaches the  $\overline{k}+1$ level at some point before maturity and ends at some level $k$ with $- \overline{l} \leq k \leq \overline{k}$, and  we define its reflection as the path that behaves like the original path up until the first time the original path touches the $\overline{k}+1$ level, and afterwards has a +1 jump when the other has a -1 jump and viceversa. Time intervals with no jump for the original path are intervals where the reflection has no jump too. The reflection path will end up at $2\overline{k}-k+2$.
%
	

Both the original path's and the reflection's probabilities are not greater then the value obtained by substituting both $\qu$ and $\qmu$ with $\frac{w}{n}$, and the sum over all paths reaching $k$ surpassing $\overline{k}$ of these modified probabilities is $\qtju (2\overline{k}+2-k) $,  therefore we can
write
\begin{equation*}\qju^{\overline{k}} (k)  \leq \qtju (2\overline{k}+2-k).\end{equation*}
Similarly, the number of paths that reach the $-\overline{l}-1$ level at some point before maturity and end at a level $k$ with $-\overline{l} \leq k \leq \overline{k}$ is the same as the number of paths that end at level $-2 \overline{l}-2-k$, and both the probability of the original path and that of its reflection with respect to the level $-\overline{l}-1$ are not greater than the modified ones, therefore it holds:
\begin{equation*}{\qju}_{\overline{l}} (k)  \leq \qtju (-2\overline{l}-2-k) = \qtju(2 \overline{l} +k+2).\end{equation*}
\end{pf}

The following Proposition allows us to find an upper estimate for the probability of reaching the upper and lower leaves of the tree, and hence for their contribution to the evaluation.

\begin{proposition}
\label{boundqtju}
For integers $k, \overline{k} \leq n $ 
\begin{align}
 \mbox{For } k \geq 2w -1 & \qquad  \qtju (k) \leq 2 \frac{w^k}{k!}  e^{w}.\\
 \label{aggiuntoqtju} \mbox{For } \overline{k} \geq 2 w -1 & \qquad 
 \sum_{k=\overline{k}}^{n}  \qtju(k)  \leq  4  e^{w}  \frac{w^{\overline{k}}}{\overline{k}!} \\
\label{eqsommasu} \mbox{ For } \overline{k} \geq 2 e^h w -1 & \qquad  
 \sum_{k=\overline{k}}^{n}  e^{hk} \qtju(k)  \leq  4  e^{w}  \frac{(e^h w)^{\overline{k}}}{\overline{k}!} \\
\label{eqsommagiu} \mbox{ For } \overline{k} \geq 2 w -1 & \qquad  
  \sum_{k=\overline{k} }^{n}  e^{-hk} \qtju(-k) \leq  4  e^{w}  \frac{(e^{-h} w)^{\overline{k}}}{\overline{k}!} 
\end{align}
\end{proposition}

\begin{pf}
Recall that \begin{equation*} \qtju (k) = \sum_{t=0}^{\left \lfloor \frac{n-k}{2} \right \rfloor}\frac{n!}{(n-k-2t)!(k+t)!t!} \left( \frac{w}{n} \right)^{k+2t} q_0^{n-k-2t}. \end{equation*}

Since $q_0 < 1$ and $ \displaystyle\frac{n!}{(n-k-2t)!n^{k+2t}} \leq 1$ for $ t = 0, \ldots, \left \lfloor \frac{n-k}{2} \right \rfloor$,  we can write:
\begin{align*}
\qtju(k) &  \leq \sum_{t=0}^{\left \lfloor \frac{n-k}{2} \right \rfloor} \frac{w^{k+2t}}{(k+t)!t!}   \\
& \leq  \frac{w^{k}}{k!}+\frac{ w^{k+1} w }{(k+1)!}+\frac{w^{k+2} w^2}{(k+2)! 2!}+...+\frac{w ^{\left \lfloor \frac{n+k}{2} \right \rfloor}w ^{\left \lfloor \frac{n-k}{2} \right \rfloor}}{\left \lfloor \frac{n+k}{2} \right \rfloor! \left \lfloor \frac{n-k}{2} \right \rfloor}  \\
& \leq \left( \frac{w^{k}}{k!}+\frac{ w^{k+1}}{(k+1)!}+\frac{w^{k+2}}{(k+2)!}+...+\frac{w^{\left \lfloor \frac{n+k}{2} \right \rfloor}}{\left \lfloor \frac{n+k}{2} \right \rfloor!} \right) \left( 1+w+\frac{w^2}{2}+...+\frac{w^{ \left \lfloor \frac{n-k}{2} \right \rfloor  }}{ \left \lfloor  \frac{n-k}{2}  \right \rfloor !} \right) \\
& \leq  \sum_{i = k}^{  \left \lfloor \frac{n+k}{2} \right \rfloor } \frac{w^i}{i!} e^{w} \leq
2 e^{w} \frac{w^k}{k!}
\end{align*}
by Lemma \ref{resto} with $j=2$, for $0 \leq w \leq \frac{k+1}{2}$.

Applying again Lemma \ref{resto} we obtain Eq. (\ref{aggiuntoqtju}).

A further application of Lemma \ref{resto} with $x=e^h w \leq \frac{\overline{k}+1}{2}$
gives us:
\begin{equation*} \sum_{k=\overline{k}}^{n}  e^{hk} \qju(k) \leq  \sum_{k=\overline{k}}^{n}  e^{hk} \qtju(k) \leq 2 e^{ w }   \sum_{k=\overline{k}}^{n}   \frac{(e^h w )^{k}}{k!} \leq  4  e^{ w }  \frac{(e^h w )^{\overline{k}}}{\overline{k}!}\end{equation*}
 for $\overline{k} \geq 2 e^h w -1 $.

Similarly for $\overline{k} \geq 2 w -1   $
we obtain Eq. (\ref{eqsommagiu}).
\end{pf}

Now all pieces are in place and we can prove the main theorems in the case $\nj =1$.

\subsubsection{Call case}

Applying  Lemma \ref{rientro1} to Eq. \ref{tagliodentro} with $\nj=1$ we can write:
\begin{align} \notag V^T - V^{TT} & \leq   e^{(\alpha-r)\tau}  S_0  \sum_{k=-\overline{l} }^{\overline{k}} e^{hk}  (\qtju (2\overline{k}+2-k) + \qtju (-2\overline{l}-2-k))  \\
 \label{perdopo} & \leq  e^{(\alpha-r)\tau}  S_0  \left(\sum_{s=\overline{k} +2 }^{\min\{2\overline{k}+\overline{l},n\}} e^{h(2\overline{k}+2-s)} \qtju (s) + \sum_{s=\overline{l} +2 }^{\min\{2\overline{l}+\overline{k},n\}} e^{h(s-2\overline{l}-2)} \qtju (s) \right)  \\
\label{tagliodentro3} & \leq  e^{(\alpha-r)\tau}  S_0  e^{h\overline{k}}  \left(\sum_{s=\overline{k} +2 }^{\min\{2\overline{k}+\overline{l},n \}} \qtju (s) + \sum_{s=\overline{l} +2 }^{\min \{2\overline{l}+\overline{k},n\}} \qtju (s) \right)
\end{align}
Combining Eq. (\ref{tagliofuori}) with $\nj=1$ and  (\ref{tagliodentro3}) and supposing we take $\overline{l}=\overline{k}$, we obtain:
\begin{equation}
\label{macro2}  V - V^{TT}  \leq 2 e^{(\alpha-r)\tau}  S_0 \left( \sum_{k = \overline{k}+1}^n (e^{hk} +  e^{h\overline{k}} )  \qtju(k)   \right).
\end{equation}
We can compute  (\ref{macro2}) with a $O(n^2)$ procedure,
 thus determining numerically the largest integer $\overline{l}=\overline{k}$ such that the loss is  inferior to an arbitrary $\varepsilon$.

Proposition \ref{boundqtju} allows us to find a theoretical bound for $\overline{k}$ and $\overline{l}$ such that   $V - V^{TT}$ is inferior to an arbitrary $\varepsilon$, and show that - using these theoretical bounds - the procedure is further reduced in complexity.

\begin{theorem} \label{errorback}
Given $\varepsilon >0 $, for $\overline{l} \geq \max \{  - \ln \varepsilon + we^{-h+1} + \ln (2+e^{h}w) + c,  2w -2, 2e^hw-3 \}$ and $\overline{k} \geq \max \{ - \ln \varepsilon + we^{h+1}+ \ln (2+e^{-h}w) +c , 2e^h w -2 \}$ with
$ c= w + (\alpha -r) \tau  -1 + \ln (4S_0) $
 we have
\begin{equation*} V - V^{TT}  < \varepsilon\end{equation*}
\end{theorem}

\begin{pf}
Combining Eq. (\ref{tagliofuori}) and  (\ref{perdopo}) with $\nj =1$ we can write:
\begin{align}
 V - V^{TT} & \leq e^{(\alpha-r)\tau}  S_0 \left( \sum_{k= \overline{k}+1}^n e^{hk} \qtju(k)  + \sum_{k = \overline{l}+1}^n  e^{-hk}  \qtju(k)  +  e^{h(2\overline{k}+2)} \sum_{k=\overline{k} +2 }^n e^{-hk} \qtju (k) + e^{h(-2\overline{l}-2)} \sum_{k=\overline{l} +2 }^n e^{hk} \qtju (k) \right)
\end{align}

\noindent By Proposition  \ref{boundqtju} we obtain:
\begin{align}
 V - V^{TT} & \leq e^{(\alpha-r)\tau}  S_0 \left( 4 e^w  \frac{(e^h w)^{\overline{k}+1}}{(\overline{k}+1)!} + 4 e^w  \frac{(e^{-h} w)^{\overline{l}+1}}{(\overline{l}+1)!} +  e^{h(2\overline{k}+2)} \cdot 4 e^w  \frac{(e^{-h} w)^{\overline{k}+2}}{(\overline{k}+2)!} + e^{h(-2\overline{l}-2)} \cdot  4 e^w  \frac{(e^h w)^{\overline{l}+2}}{(\overline{l}+2)!}  \right) \\
& \leq 4 e^{(\alpha-r)\tau+w}  S_0 \left(  \frac{(e^h w)^{\overline{k}+1}}{(\overline{k}+1)!} +  \frac{(e^{-h} w)^{\overline{l}+1}}{(\overline{l}+1)!} +  e^{-2h} \cdot   \frac{(e^{h} w)^{\overline{k}+2}}{(\overline{k}+2)!} + e^{2h} \cdot  \frac{(e^{-h} w)^{\overline{l}+2}}{(\overline{l}+2)!}  \right) \\
& \leq 4 e^{(\alpha-r)\tau+w}  S_0 \left( \left(1+ \frac{e^{-h} w}{\overline{k}+2}\right) \frac{(e^h w)^{\overline{k}+1}}{(\overline{k}+1)!} + \left(1+ \frac{e^{h} w}{\overline{l}+2}\right)   \frac{(e^{-h} w)^{\overline{l}+1}}{(\overline{l}+1)!} \right) \\
& \leq 4 e^{(\alpha-r)\tau+w}  S_0 \left( \left(1+ \frac{e^{-h} w}{2}\right) \frac{(e^h w)^{\overline{k}+1}}{(\overline{k}+1)!} + \left(1+ \frac{e^{h} w}{2}\right)   \frac{(e^{-h} w)^{\overline{l}+1}}{(\overline{l}+1)!} \right)
\end{align}
\noindent for $ \overline{k} \geq 2 e^h w -2$ and $ \overline{l}\geq \max \{ 2 e^h w -3, 2w-2 \}$

If we equally split the error $\varepsilon$ between the upper and the lower region, we ask:
\begin{align}
 \label{lower} 4 S_0 e^{(\alpha -r) \tau  +w}  \left( 1+   \frac{e^{h} w }{2} \right) \frac{(e^{-h} w)^{\overline{l}+1}}{(\overline{l}+1)!} &  < \frac{\varepsilon}{2} \\
\label{upper} 4 S_0 e^{(\alpha -r) \tau +w}   \left( 1+  \frac{e^{-h} w }{2} \right)  \frac{(e^h w)^{\overline{k}+1}}{(\overline{k}+1)!} &  < \frac{\varepsilon}{2}
\end{align}

Taking $C = 4 S_0 e^{(\alpha -r) \tau +w} $,
 Eq. (\ref{lower}) is equivalent to
\begin{equation*} \frac{(e^{-h} w)^{\overline{l}+1}}{(\overline{l}+1)!}   < \frac{\varepsilon}{(2+ e^{h} w ) C}, \end{equation*}
which is guaranteed by Lemma \ref{logari} for $\overline{l} > - \ln \varepsilon + \ln [ (2+ e^{h} w )C ] + e^{-h+1}w -1 $, while  Eq. (\ref{upper}) is equivalent to

\begin{equation*} \frac{(e^{h} w)^{\overline{k}+1}}{(\overline{k}+1)!}  < \frac{\varepsilon}{(2+ e^{-h} w )C} \end{equation*}

which is guaranteed by Lemma \ref{logari} for $\overline{k} \geq - \ln \varepsilon + \ln [ (2+ e^{-h} w )C ] + e^{h+1}w -1 $.

\noindent Taking into account all the previous conditions, the thesis is guaranteed for

\begin{equation*}  \overline{l} \geq \max \{  - \ln \varepsilon + we^{-h+1} +w + (\alpha -r) \tau  -1 + \ln (4S_0) + \ln (2+e^{h}w),  2w -2, 2e^hw-3 \}\end{equation*}

and

\begin{equation*}  \overline{k} \geq \max \{ - \ln \varepsilon + we^{h+1}+w  + (\alpha -r) \tau  -1 + \ln (4S_0) + \ln (2+e^{-h}w) , 2e^h w -2 \}. \end{equation*}

\end{pf}

\begin{theorem} \label{complexEuro}
Given $\varepsilon=\frac{1}{n} >0 $, and $\overline{k}$ and $\overline{l}$ the smallest integers  as in Theorem \ref{errorback}, the proposed procedure for $V^{TT}$ converges to the HS price and its computational complexity is  $O(n \ln n)$. 
\end{theorem}

\begin{pf}
 By taking $\overline{k}=\overline{l}$ the smallest integer as in Theorem \ref{errorback}, the error is given by
\begin{equation*} \left \lvert V -  e^{-r\tau} \sum_{k=-\overline{l}}^{\overline{k}} \sum_{j=0}^{n} (S_0 e^{(-n+2j)\sigma \sqrt{\Delta t}+hk} - \strike)^+ \pscad(j)  \qwju (k) \right \rvert 
< \varepsilon \end{equation*}
The sum over $k$ has at most a number of terms proportional to $\ln n$, the approximate probability distribution $\qwju$ is $O(n \ln n)$, therefore the computational complexity of the procedure is $O(n \ln n)$.
\end{pf}

The previous theorem guarantees that for appropriate $\overline{k}$ and $\overline{l}$, the value $V$ of the European call option can be approximated by
\begin{equation*}V^{TT} = e^{-r\tau} \sum_{k=-\overline{l}}^{\overline{k}} \sum_{j=0}^{n} (S_0 e^{(-n+2j)\sigma \sqrt{\Delta t}+hk} - 
\strike)^+ \pscad(j)  \qwju (k)\end{equation*}

\subsubsection{Put case}

We act exactly as in the call case. Applying  Lemma \ref{rientro1} to Eq. \ref{tagliodentroput} with $\nj=1$ we can write:
\begin{align} \notag V^T - V^{TT} & \leq   e^{-r\tau}  \strike  \sum_{k=-\overline{l} }^{\overline{k}} (\qtju (2\overline{k}+2-k) + \qtju (-2\overline{l}-2-k))  \\
\label{tagliodentro3put} & \leq  e^{-r\tau}  \strike  \left(\sum_{s=\overline{k} +2 }^{\min\{2\overline{k}+\overline{l},n \}} \qtju (s) + \sum_{s=\overline{l} +2 }^{\min \{2\overline{l}+\overline{k},n\}} \qtju (s) \right)
\end{align}
Combining Eq. (\ref{tagliofuoriput}) with $\nj=1$ and  (\ref{tagliodentro3put}) and supposing we take $\overline{l}=\overline{k}$, we obtain:
\begin{equation}
\label{macro2put}  V - V^{TT}  \leq 4 e^{-r \tau} \strike \left( \sum_{k = \overline{k}+1}^n \qtju(k)   \right).
\end{equation}
We can compute  (\ref{macro2put}) such that   $V - V^{TT}$ is inferior to an arbitrary $\varepsilon$ with a $O(n^2)$ procedure, but a theoretical bound for $\overline{k}$ and $\overline{l}$ is provided by Proposition \ref{boundqtju}.

\begin{theorem} \label{errorbackput}
Given $\varepsilon >0 $, for $\overline{l}, \overline{k} \geq \max \{ - \ln \varepsilon +  c, 2 w -2 \} $ with
$ c=  w(e +1)  -r\tau  -1 + \ln (4 \strike) + \ln (2+w)$
 we have
\begin{equation*} V - V^{TT}  < \varepsilon\end{equation*}
\end{theorem}

\begin{pf}
Combining Eq. (\ref{tagliofuoriput}) with $\nj=1$ and  (\ref{tagliodentro3put})  and applying Proposition  \ref{boundqtju} we obtain:
\begin{align}
 V - V^{TT} & \leq e^{-r \tau}  \strike \left( 4 e^w  \frac{w^{\overline{k}+1}}{(\overline{k}+1)!} + 4 e^w  \frac{ w^{\overline{l}+1}}{(\overline{l}+1)!} +  4 e^w  \frac{w^{\overline{k}+2}}{(\overline{k}+2)!} + 4 e^w  \frac{w^{\overline{l}+2}}{(\overline{l}+2)!}  \right) \\
& \leq 4  e^{w-r \tau}  \strike  \left(  \frac{w^{\overline{k}+1}}{(\overline{k}+1)!} +  \frac{w^{\overline{l}+1}}{(\overline{l}+1)!} +  \frac{w^{\overline{k}+2}}{(\overline{k}+2)!} + \frac{w^{\overline{l}+2}}{(\overline{l}+2)!}  \right) \\
& \leq 4  e^{w-r \tau}  \strike  \left(1+ \frac{w}{2}\right) \left(  \frac{w^{\overline{k}+1}}{(\overline{k}+1)!} + \frac{w^{\overline{l}+1}}{(\overline{l}+1)!} \right)
\end{align}
\noindent for $ \overline{k}, \overline{l} \geq 2 w -2$.
If we equally split the error $\varepsilon$ between the upper and the lower region, we ask $\overline{k}=\overline{l}$ such that:
\begin{equation}
 \label{lowerupper} 4  e^{w-r \tau}  \strike  \left(1+ \frac{w}{2}\right)  \frac{w^{\overline{k}+1}}{(\overline{k}+1)!}  < \frac{\varepsilon}{2} 
\end{equation}
Taking $C = 4  e^{w-r \tau}  \strike $,
 Eq. (\ref{lowerupper}) is equivalent to
\begin{equation*} \frac{w^{\overline{k}+1}}{(\overline{k}+1)!}   < \frac{\varepsilon}{(2+w)C}, \end{equation*}
which is guaranteed by Lemma \ref{logari} for $\overline{k} > - \ln \varepsilon + \ln [ (2+ w )C ] + ew -1 $.
\noindent Taking into account all the previous conditions, the thesis is guaranteed for
\begin{equation*}  \overline{l},\overline{k} \geq \max \{ - \ln \varepsilon + w(e+1)  -r\tau  -1 + \ln (4 \strike) + \ln (2+w) , 2w -2 \}. \end{equation*}
\end{pf}

Then we can state also in the put case the analogous to Theorem \ref{complexEuro}:

\begin{theorem} \label{complexEuroput}
Given $\varepsilon=\frac{1}{n} >0 $, and $\overline{k}=\overline{l}$ the smallest integer  as in Theorem \ref{errorbackput}, the proposed procedure for $V^{TT}$ converges to the HS price and its computational complexity is  $O(n \ln n)$. 
\end{theorem}

\subsection{European option, arbitrary $\nj$}

We extend the results of the previous Section to arbitrary $\nj$. For brevity, proofs are available in the online appendix.
The analogous to Lemma \ref{rientro1} and Proposition \ref{boundqtju} are the following.

\begin{lemma} \label{rientronj}
\begin{equation*}\qjn^{\overline{k}}(k) \leq \sum_{i=1}^{\nj} \qtjn (2 \overline{k} -k+2i) \end{equation*}
\begin{equation*}{\qjn}_{\overline{l}}(k) \leq \sum_{i=1}^{\nj} \qtjn (2 \overline{l} +k+2i) \end{equation*}
for all $-\overline{l} \leq k \leq \overline{k}$.
\end{lemma}

\begin{proposition}  \label{limitinj}
For integers $k, \overline{k} \leq \nj n$
\begin{align}
\label{3sint} \mbox{ For } k \geq \nj \lceil 2 W_{\nj}-1 \rceil & \qquad   \qtjn( k)  \leq
 G \frac{W_{\nj}^{\left \lfloor \frac{k}{\nj} \right \rfloor}}{\left \lfloor \frac{k}{\nj} \right \rfloor!} \\
\label{perputn} \mbox{ For } \overline{k} \geq  \nj  \lceil 2 W_{\nj} - 1 \rceil & \qquad    \sum_{k=\overline{k}}^{\nj n}   \qtjn(k) \leq
2 G \nj  \frac{W_{\nj}^{\left \lfloor \frac{\overline{k}}{\nj} \right \rfloor}}{ \left \lfloor \frac{\overline{k}}{\nj} \right \rfloor !}
\\
\mbox{ For } \overline{k} \geq  \nj  \lceil 2 e^{\nj h} W_{\nj} - 1 \rceil & \qquad  \sum_{k=\overline{k}}^{\nj n}  e^{hk} \qtjn(k) \leq
2 G  \frac{(e^{h \nj} W_{\nj})^{\left \lfloor \frac{\overline{k}}{\nj} \right \rfloor}}{ \left \lfloor \frac{\overline{k}}{\nj} \right \rfloor !} \sum_{r=0}^{\nj-1}  e^{hr} \\
\mbox{ For } \overline{k} \geq \nj  \lceil 2 W_{\nj} - 1 \rceil & \qquad 
\sum_{k=\overline{k}}^{\nj n}  e^{-hk} \qtjn(-k) \leq
2 G \frac{(e^{-h \nj} W_{\nj})^{\left \lfloor \frac{\overline{k}}{\nj} \right \rfloor}}{ \left \lfloor \frac{\overline{k}}{\nj} \right \rfloor !} \sum_{r=0}^{\nj-1}  e^{-hr}
\end{align}
\end{proposition}

\subsubsection{Call case}

We take $V$, $V^T$ and   $V^{TT}$ as defined in Eq. (\ref{euroV}), (\ref{HSN}) and (\ref{vethat}) respectively.

By Lemma \ref{rientronj} we have
\begin{align} \notag V^T - V^{TT} & \leq   e^{(\alpha-r)\tau}  S_0   \left( \sum_{k=-\overline{l} }^{\overline{k}} e^{hk}   \sum_{i=1}^{\nj} \qtjn (2 \overline{k} -k+2i)  + \sum_{k=-\overline{l} }^{\overline{k}} e^{hk}  \sum_{i=1}^{\nj} \qtjn (+2 \overline{l} +k+2i) \right)   \\
\label{perdopon} & \leq   e^{(\alpha-r)\tau}  S_0  \left( \sum_{s=\overline{k} + 2}^{\min\{2\overline{k}+\overline{l}+2, \nj n \}}  e^{h(2\overline{k}-s+2)}  \sum_{i=0}^{\nj-1} \qtjn (s +2i) + \sum_{s=\overline{l} +2 }^{\min \{2\overline{l}+\overline{k}+2, \nj n\}} e^{h(s-2\overline{l}-2)}  \sum_{i=0}^{\nj-1} \qtjn (s+2i) \right)  \\
\label{tagliodentron} & \leq   e^{(\alpha-r)\tau}  S_0  e^{h\overline{k}} \nj \left( \sum_{s=\overline{k} + 2}^{ \nj n }    \qtjn (s) + \sum_{s=\overline{l} +2 }^{ \nj n } \qtjn (s) \right)
\end{align}
Combining Eq. (\ref{tagliofuori}) and  (\ref{tagliodentron}) we obtain:
\begin{align}
V - V^{TT} & \leq  e^{(\alpha-r)\tau}  S_0 \left( \sum_{k=\overline{k}+1}^{ \nj n}  (e^{hk} +  e^{h\overline{k}} \nj ) \qtjn(k)  + \sum_{k=\overline{l}+1}^{ \nj n}  ( e^{-hk}   + e^{h\overline{k}} \nj )  \qtjn(k)  \right)  
\end{align}

The closed formulas for $\overline{l}$ and $\overline{k}$ specified in Theorem \ref{errorbacknj} can be retrieved in the same way as in the $\nj =1$ case (see appendix).

\subsubsection{Put case}

We take $V$, $V^T$ and   $V^{TT}$ the values for the put option analogous to those previously defined for the call.

By Lemma \ref{rientronj} applied to Equation  \ref{tagliodentroput} we have
\begin{align} \notag V^T - V^{TT} & \leq   e^{-r\tau} K  \left( \sum_{k=-\overline{l} }^{\overline{k}} \sum_{i=1}^{\nj} \qtjn (2 \overline{k} -k+2i)  + \sum_{k=-\overline{l} }^{\overline{k}} \sum_{i=1}^{\nj} \qtjn (+2 \overline{l} +k+2i) \right)   \\
\label{tagliodentronput} & \leq   e^{-r\tau}  K  \nj \left( \sum_{s=\overline{k} + 2}^{ \nj n }    \qtjn (s) + \sum_{s=\overline{l} +2 }^{ \nj n } \qtjn (s) \right)
\end{align}
Combining Eq. (\ref{tagliofuoriput}) and  (\ref{tagliodentronput}) we obtain:
\begin{align}
V - V^{TT} & \leq  e^{-r\tau} K (\nj +1) \left( \sum_{k=\overline{k}+1}^{ \nj n}  \qtjn(k)  + \sum_{k=\overline{l}+1}^{ \nj n}  \qtjn(k)  \right)  
\end{align}

The closed formula for $\overline{l}=\overline{k}$ specified in Theorem \ref{putnj} can be retrieved in the same way as in the $\nj =1$ case (see appendix).

\section{American option}

In the following, we extend the results on the backward procedure for the evaluation of the European derivatives to the American put option pricing, by showing that the truncation error for the American case must be less or equal then the error in the European case.

Recall the definition of $V_E^b(0,0,0)$ and $V_A^b(0,0,0)$, which are the prices 
 we obtain via backward recursion on a tree where the value of the option in any node outside the allowed region is substituted by $b \geq 0$. We remark that $V_E^0(0,0,0)$ corresponds to $V^{TT}$. 

\begin{lemma}
$V_E^0(0,0,0)=V^{TT}$ 
\end{lemma}

The simple proof is available in the Appendix for the reader's convenience.

We also consider, both in the call and in the put case, the value $\widehat{V^b}$, defined as

\begin{equation}
\label{VEK}
 	\widehat{V^b} = V^{TT} + \sum_{\mbox{paths that reach $\tau$ and trespass}} \mbox{prob(path)} \cdot b e^{-r \Delta t \mbox{i(path)}} 
	\end{equation}

where prob(path) identifies the probability of a single path and i(path) the time $0 <i \leq n$ of the first exit of the path from the allowed zone.

One can easily show (proof in the appendix) that the value $V_E^b(0,0,0)$ coincides with $\widehat{V^b}$:
\begin{lemma}
$V_E^b(0,0,0)=\widehat{V^b}$ 
\end{lemma}

We consider $b=K$.

Substituting $0$ with $\strike$ in the nodes above the barrier increases the value of the option, therefore we have $V_E^K (0,0,0) \geq V_E^0(0,0,0)$. Since for the put options we also have $V_E(i,j,k) \leq \strike$ for every $(i,j,k)$, we have that
$V_E^K (0,0,0) \geq V_E(0,0,0) \geq V_E^0(0,0,0)$.

Therefore 

$$|V_E^K (0,0,0) - V_E(0,0,0) |\leq | V_E^K (0,0,0) - V_E^0(0,0,0)|  = | \widehat{V^K} - V^{TT} |.$$

In order to control $|V_E^K (0,0,0) - V_E(0,0,0) |$ we only need to control $\widehat{V^K} - V^{TT}  $.

\begin{lemma} \label{bordoamenotte}
Given $\varepsilon >0 $, taking $G =  2 \nj \max \{ W_{\nj}, 1 \}  \prod_{i= 1}^{\nj -1} M^2_i   e^{W_{\nj}} $, the values $\widehat{V^K}$ and $V^{TT}$ obtained via truncation of the tree at levels  $\overline{k}$ and $-\overline{k}$, with $\overline{k}$ the smallest integer which satisfies:

$\overline{k} \geq \max \{  \nj  \lceil 2 W_{\nj} - 1 \rceil -1 ,  \nj  \lceil W_{\nj} e -\ln \varepsilon  + \ln(4 \nj  (\nj+1) \strike G)  \rceil -1 \}$, we have

\begin{equation*} \left \lvert \widehat{V^K} - V^{TT} \right \rvert < \varepsilon\end{equation*}
\end{lemma}

Proof of Lemma \ref{bordoamenotte} is given in the Appendix, here we will show the proof of Theorem \ref{American}.

\begin{pfnome} \textbf{of Theorem \ref{American}}

Consider $V_E^K(i,j,l)$, $V_A^K(i,j,l)$, $i=0,..., n$, the backward procedure which has value $K$ under the $-\overline{l}$ and over the $\overline{k}$ barrier for all the time steps $i$, running from $0$ to $n$ and $V_E(i,j,l)$, $V_A(i,j,l)$, $i=0,...,n$ the standard backward procedure, as defined at page \pageref{web}.

We claim that $$|V_A^K(i,j,l)-V_A(i,j,l)|\leq |V_E^K(i,j,l)-V_E(i,j,l)|$$ for all $i,j,l$.

Since the American put price is not smaller than the European price and it is not larger than $K$, out of the barriers the claim is true. Inside the boundaries, we prove it for backward induction on the timestep $i$.

Let $i=n$. On all the nodes at maturity the error is the same, since $V_A^K(n,j,l)=V_E^K(n,j,l)$ and $V_A(n,j,l)=V_E(n,j,l)$.

Consider now the case $i-1$. 

We set the continuation value $C_A^K(i-1,j,l) =e^{-r\Delta t} \sum_{k = -\nj}^{\nj} (V_A^K(i, j+1, l+k) p + V_A^K(i, j, l+k) (1-p)) q_k$ and $C_A(i-1,j,l)$, $C_E^K(i-1,j,l)$, $C_E(i-1,j,l)$, similarly. Consider the nodes inside the barriers.  The truncation value $V_A^K(i-1,j,l)$ is then given by

\begin{equation}
 V_A^K(i-1,j,l)=\max \left[ C_A^K(i-1,j,l) , K-S(i-1,j,l) \right]
\end{equation}

One has $C_A(i,j,l) \leq C_A^K(i,j,l)$ for every $i, j, l$.

Only the following cases are possible:

\begin{itemize}
\item $C_A^K(i-1,j,l) \leq K-S(i-1,j,l)$

 This means $V_A^K(i-1,j,l)=K-S(i-1,j,l)=V_A(i-1,j,l)$, and  $|V_A^K(i-1,j,l)-V_A(i-1,j,l)| = 0$.

\item $C_A(i-1,j,l) \leq K-S(i-1,j,l)$ and $C_A^K(i-1,j,l) \geq K-S(i-1,j,l)$

  This means $V_A(i-1,j,l)=K-S(i-1,j,l)$ and $V_A^K(i-1,j,l)=C_A^K(i-1,j,l)$, and
	
	\begin{align*}
	& |V_A^K(i-1,j,l)-V_A(i-1,j,l)|  = C_A^K(i-1,j,l) - (K-S(i-1,j,l)) \\
	& \leq C_A^K(i-1,j,l) - C_A(i-1,j,l) \\
	\leq &  e^{-r\Delta t}  \sum_{k = -\nj}^{\nj} (V_A^K(i, j+1, l+k) p + V_A^K(i, j, l+k) (1-p)) q_k + \\
	& -   e^{-r\Delta t}  \sum_{k = -\nj}^{\nj} (V_A(i, j+1, l+k) p + V_A(i, j, l+k) (1-p)) q_k \\
	=  & e^{-r\Delta t} \sum_{k = -\nj}^{\nj} [(V_A^K(i, j+1, l+k)-V_A(i, j+1, l+k)] p q_k + \\
&	+ e^{-r\Delta t} \sum_{k = -\nj}^{\nj} [V_A^K(i, j, l+k)- V_A(i, j, l+k)] (1-p) q_k 
	\end{align*}

Either the nodes $(i, j+1, l+k)$ and $(i, j, l+k)$ are outside the boundaries, and then the claim is true, or we can use induction, therefore:  $V_A^K(i, j+1, l+k)-V_A(i, j+1, l+k) \leq V_E^K(i, j+1, l+k)-V_E(i, j+1, l+k)$, and $V_A^K(i, j, l+k)-V_A(i, j, l+k) \leq V_E^K(i, j, l+k)-V_E(i, j, l+k)$. Hence:

	\begin{align*}
& |V_A^K(i-1,j,l)-V_A(i-1,j,l)| \leq \\
 \leq & e^{-r\Delta t} \sum_{k = -\nj}^{\nj}  [(V_E^K(i, j+1, l+k)-V_E(i, j+1, l+k)] p q_k \\
  & + e^{-r\Delta t} \sum_{k = -\nj}^{\nj}   [V_E^K(i, j, l+k)- V_E(i, j, l+k)] (1-p) q_k \\
	 \leq &  e^{-r\Delta t}  \sum_{k = -\nj}^{\nj} (V_E^K(i, j+1, l+k) p + V_E(i, j, l+k) (1-p)) q_k + \\
	& -   e^{-r\Delta t}  \sum_{k = -\nj}^{\nj} (V_E^K(n, j+1, l+k) p + V_E(i, j, l+k) (1-p)) q_k \\
	 \leq & V_E^K(i,j,l)-V_E(i,j,l).
	\end{align*}

\item $C_A(i-1,j,l) \geq K-S(i-1,j,l)$

  This means $V_A^K(i-1,j,l)=C_A^{K}(i-1,j,l)$ and $V_A(i-1,j,l)=C_A(i-1,j,l)$, and

$$|V_A(i-1,j,l)-V_A^K(i-1,j,l)| = C_A^K(i-1,j,l) - C_A(i-1,j,l) $$

which we have already considered in the previous case.
\end{itemize}
\end{pfnome}

Theorem \ref{American} and Lemma \ref{bordoamenotte} allow us to state Theorem \ref{ameback}, a result for American put options analogous to Theorem \ref{complexEuronj}.
 Proof of Theorem \ref{ameback} is straightforward.

\section{Tables} \label{numtest}

Even though we are able to specify precise theoretical values for $\overline{l}$ and $\overline{k}$, in the numerical computations for this work we have proceeded in the following way.
For the call option we have considered the following conditions:
 $\sum_{k=\overline{k}+1}^{ \nj n} (\nj+1) e^{hk} \qtjn(k) < \frac{\varepsilon}{2}$ and $ \sum_{k=\overline{l} +1 }^{ \nj n } ( \nj e^{h \overline{k}} +1) \qtjn (k) < \frac{\varepsilon}{2} $.
 Given $\varepsilon$, we consider
 $\eta= \frac{\varepsilon}{2 e^{(\alpha -r) \tau} S_0 }$ and we compute, starting from $i=\nj n$,
the sums $S^e_i = (\nj+1) \sum_{k=i}^{\nj n } e^{hk} \qtjn(k)$ and $S_i = \sum_{k=i}^{\nj n } \qtjn(k)$.
While $S^e_i < \eta$, we keep decreasing $i$. The first $i$ we encounter such that $S^e_i \geq \eta$ is our $\overline{k}$. While $S_i < \eta$, we keep decreasing $i$.  The first $i$ we encounter such that $S_i < \frac{\eta}{\nj e^{h \overline{k}} +1}$ is our $\overline{l}$.
For the put option, given $\varepsilon$, we consider
 $\eta= \frac{\varepsilon}{2 e^{-r\tau}  \strike (\nj+1)  }$ and we compute, starting from $i=\nj n$,
the sum $S_i = \sum_{k=i}^{\nj n } \qtjn(k)$. While $S_i < \eta$, we keep decreasing $i$.  The first $i$ we encounter such that $S_i \geq \eta$ is our $\overline{l} = \overline{k}$.

Determining numerically in this way the largest integers $\overline{l}$ and $\overline{k}$ such that the loss is  inferior to an arbitrary $\varepsilon$ is an $O(n^2)$ procedure, therefore the American procedure is still $O(n^2 \ln n)$.

We compare our results with the ones obtained by the procedures described by Hilliard and Schwartz \cite{HS}, Amin \cite{Amin}, and Dai \textit{et al.} \cite{Lyuu}, and closed formula by Merton, reporting the calculation times 
 in order to highlight the advantage provided by our procedures. 

Table 1 and 2 compare the European put option prices obtained with different methods and the corresponding computational times, for several strikes, using as a benchmark the value from Merton series formula. In Table 1 different number of steps are used, with constant time to maturity $\tau=1$ year. In Table 2 the procedures are compared in different maturities and variances $\sigma^2$, while $n=400$, $\gamma=0$, $\delta^2=0.05$.
In all calculations $\nj=3$ has been used in Dai, HS, and our procedure; $r=0.08$, $d=0.00$, Poisson parameter $\lambda=5.0$, current value $S_0=40$ in both Tables.

We may note that Dai \textit{et al.} method is sensitive to the value of $\sigma$, while our method is sensitive to the increasing time to maturity.

In Tables 3 and 4 we compare our results with those of Simonato \cite{Simonato2011}, Amin \cite{Amin}, Hilliard and Schwartz \cite{HS} and Dai \textit{et al.} \cite{Lyuu}, for European and American call options. The benchmark for the American case in Table 4 is the price obtained via Kim's integral equation as reported in Simonato \cite{Simonato2011}. We see that the numerical results for the American call options present the same precision as in the case of put option, even in the absence of a theoretical result.

\footnotesize

\begin{tabular}{llrrrrr}
\multicolumn{7}{l}{Table 1}  \\
\toprule
Strike & Steps &  \multicolumn{5}{l}{European puts}  \\
\cmidrule(lr){3-7}
 & & \multicolumn{1}{l}{Amin} &  \multicolumn{1}{l}{Dai} & \multicolumn{1}{l}{HS}  &  \multicolumn{1}{l}{HScut} &  \multicolumn{1}{l}{Merton} \\
\midrule
 \multicolumn{7}{l}{Panel A: $\gamma=0$, $\delta^2=0.05$, $\sigma^2=0.05$} \\
& 200 & 2.6253 (0.22) &  2.6207 (0.47) & 2.6215 (1.01) &  2.6215 (0.08) \\
30 & 400 &  2.6233 (1.70) & 2.6209 (2.70) & 2.6217 (8.04) & 2.6217 (0.28) \\
& 800 &  2.6223 (13.5) & 2.6210 (14.1) & 2.6213 (63.9) &  2.6213 (1.05) & 2.6211\\
\cmidrule(lr){2-7}
& 200 &  6.7102 (0.22) &  6.6972 (0.46) & 6.6982 (1.01) & 6.6982 (0.09) \\
40 & 400 &  6.7029 (1.70) &  6.6976 (2.70) & 6.6070 (7.98) &  6.6970 (0.29) \\
& 800 &  6.6995 (13.5) & 6.6964 (14.1) & 6.6968 (63.8) &  6.6968 (1.05) & 6.6970\\
\cmidrule(lr){2-7}
& 200 &  12.5486 (0.23) & 12.5247 (0.45) & 12.5260 (1.00) &  12.5260 (0.09) \\
50 & 400 &  12.5360 (1.69) &  12.5243 (2.66) &  12.5249 (7.88) &  12.5249 (0.29) \\
& 800 &  12.5301 (13.3) & 12.5241 (13.9) &  12.5247 (61.2) &  12.5247 (1.03) & 12.5238 \\
\midrule
\multicolumn{7}{l}{Panel B: $\gamma=0$, $\delta^2=0.09$, $\sigma^2=0.01$} \\
& 200 & 3.7542 (0.20) & 3.9151 (1.24) &  3.9154 (0.98) &  3.9154 (0.08) \\
30 & 400 & 3.9086 (1.79) & 3.9138 (7.66) & 3.9141 (8.03) & 3.9141 (0.30) \\
& 800 & 3.9220 (13.8) &  3.9131 (41.2) & 3.9132 (62.4) & 3.9132 (1.09)& 3.9184 \\
\cmidrule(lr){2-7}
& 200 &  8.3061 (0.22) & 8.4652 (1.23) & 8.4654 (0.98) & 8.4654 (0.09) \\
40 & 400 & 8.4547 (1.69) & 8.4620 (7.59) &  8.4621 (8.01) &  8.4621 (0.30) \\
& 800 & 8.4648 (13.3) & 8.4603 (41.3) & 8.4604 (61.8) & 8.4604 (1.10) & 8.4578 \\
\cmidrule(lr){2-7}
& 200 & 14.3182 (0.22) & 14.4825 (1.24) & 14.4831(0.96) & 14.4831 (0.09) \\
50 & 400 & 14.4621 (1.69) & 14.4793 (7.61) & 14.4795 (7.88) & 14.4795 (0.30) \\
& 800 & 14.4697 (13.3) & 14.4778 (41.4) & 14.4778 (61.9) & 14.4778 (1.11) & 14.4604 \\
\midrule
\multicolumn{7}{l}{Panel C: $\gamma=0$, $\delta^2=0.05$, $\sigma^2=0.0025$} \\
& 200 & 1.4498 (0.23) & 2.1887 (1.82) & 2.1888 (0.96) & 2.1888 (0.09) \\
30 & 400 & 1.9766 (1.68) &  2.1883 (11.3) & 2.1884 (7.94) & 2.1884 (0.29) \\
& 800 & 2.1502 (13.3) & 2.1881 (60.6) & 2.1881 (62.0) &  2.1881 (1.05)  & 2.1720 \\
\cmidrule(lr){2-7}
& 200 &  5.2298 (0.24) & 6.0039 (1.83) & 6.0040 (0.97) & 6.0040 (0.10) \\
40 & 400 & 5.7905 (1.68) & 6.0014 (11.3) & 6.0015 (7.94) & 6.0015 (0.29) \\
& 800  & 5.9625 (13.3) & 6.0014 (60.6) & 6.0002 (62.2) & 6.0002 (1.03) & 5.9800  \\
\cmidrule(lr){2-7}
& 200 & 11.0203 (0.23) & 11.7862 (1.82) & 11.7866 (0.98) & 11.7866 (0.09) \\
50 & 400 &  11.5728 (1.68) & 11.7839 (11.3) & 11.7841 (8.01) & 11.7841 (0.28) \\
& 800 & 11.7414 (13.3) & 11.7828 (60.6) & 11.7829 (62.1) & 11.7829 (1.05) & 11.7556 \\
\bottomrule
\end{tabular}

\begin{tabular}{lrrrr}
\multicolumn{5}{l}{Table 2}  \\
\toprule
 Strike &  \multicolumn{4}{l}{European puts}  \\
\cmidrule(lr){2-5}
 &  \multicolumn{1}{l}{Amin} &  \multicolumn{1}{l}{Dai} &  \multicolumn{1}{l}{HScut} &  \multicolumn{1}{l}{Merton} \\
\midrule
\multicolumn{5}{l}{Panel A: Maturity $\tau=$ one year, $\sigma^2=0.05$} \\
30 &  2.6233 (1.70) & 2.6209 (2.70) & 2.6217 (0.28) &  2.6211 \\
40 &  6.7029 (1.70) &  6.6976 (2.70) &  6.6970 (0.29) & 6.6970 \\
50 &  12.5360 (1.69) &  12.5243 (2.66)  &  12.5249 (0.29) & 12.5238 \\
\cmidrule(lr){2-5}
\multicolumn{5}{l}{Panel B: Maturity $\tau=$ one year, $\sigma^2=0.01$} \\
30 &  2.2486 (1.68) & 2.2448 (5.72) & 2.2451 (0.27) &  2.2436 \\
40 &  6.1124 (1.68) &  6.1029 (5.72) &  6.1032 (0.27) & 6.0995 \\
50 &  11.9013 (1.68) &  11.8860 (5.72)  &  11.8864 (0.27) & 11.8819 \\
\cmidrule(lr){2-5}
\multicolumn{5}{l}{Panel C: Maturity $\tau=5$ years, $\sigma^2=0.05$} \\
30 &  5.6850 (1.68) & 5.6178 (1.28) & 5.6200 (0.50) &  5.6013 \\
40 &  9.5178 (1.68) & 9.4120 (1.28) & 9.4143 (0.50) & 9.3861 \\
50 &  13.8861 (1.68) &  13.7415 (1.27)  &  13.7446 (0.50) & 13.7055 \\
\cmidrule(lr){2-5}
\multicolumn{5}{l}{Panel D: Maturity $\tau=5$ years, $\sigma^2=0.01$} \\
30 &  5.0466 (1.68) & 4.9361 (2.64) & 4.9374 (0.50) &  4.9198 \\
40 &  8.6917 (1.68) & 8.5266 (2.64) & 8.5281 (0.50) & 8.5003 \\
50 &  12.9203 (1.68) &  12.7024 (2.64)  &  12.7042 (0.50) & 12.6657 \\
\cmidrule(lr){2-5}
\multicolumn{5}{l}{Panel E: Maturity $\tau=10$ years, $\sigma^2=0.05$} \\
30 & 5.4517 (1.68) & 5.2829 (0.93) & 5.2857 (0.73) & 5.2834 \\
40 & 8.3314 (1.68) & 8.0925 (0.93)  & 8.0972 (0.73) & 8.0927 \\
50 & 11.4521 (1.68) & 11.1450 (0.94) & 11.1495 (0.73) & 11.1450 \\
\cmidrule(lr){2-5}
\multicolumn{5}{l}{Panel F: Maturity $\tau=10$ years, $\sigma^2=0.01$} \\
30 & 4.9085 (1.68) & 4.6494 (1.89) & 4.6516 (0.73) & 4.6491 \\
40 & 7.6451 (1.68) & 7.2843 (1.89) & 7.2874 (0.73) & 7.2832 \\
50 & 10.6468 (1.68) & 10.1889 (1.89) & 10.1926 (0.73) & 10.1872 \\
\bottomrule
\end{tabular}

\begin{tabular}{lrrrrr} 
\multicolumn{6}{l}{Table 3}  \\
\toprule
 Maturity &  \multicolumn{5}{l}{European calls} \\
\multicolumn{6}{l}{$\gamma'=-0.02$, $\delta^2=0.01 $, $r=0.05$, $d=0$, $\sigma^2=0.04$, $\lambda=5$, $n=150$}  \\
\cmidrule(lr){1-6}
 &  \multicolumn{1}{l}{Simonato} &  \multicolumn{1}{l}{Amin} &  \multicolumn{1}{l}{Dai} &  \multicolumn{1}{l}{HScut}  &  \multicolumn{1}{l}{Merton} \\
\midrule
\multicolumn{6}{l}{Panel A: $K =45$ } \\
30/365 & 5.4304 & 5.4429 & 5.4430 & 5.4435  & 5.4582 \\
90/365 & 6.4372 & 6.4263 & 6.4367 & 6.4389  & 6.4607 \\
270/365 & 8.8432 & 8.7390 & 8.8323 & 8.8362 & 8.8668 \\
\cmidrule(lr){2-6}
\multicolumn{6}{l}{Panel B: $K =50$ } \\
30/365 & 1.7306 & 1.6952 & 1.6960 & 1.6961 & 1.7038 \\
90/365 & 3.2149 & 3.1879 & 3.1952 & 3.1964  & 3.2119 \\
270/365 & 5.9859 & 5.8932 & 5.9731 & 5.9773  & 6.0041 \\
\cmidrule(lr){2-6}
\multicolumn{6}{l}{Panel C: $K =55$ } \\
30/365 & 0.3030 & 0.3026 & 0.3023 & 0.3031 & 0.2936 \\
90/365 & 1.3251 & 1.3111 & 1.3152 & 1.3176  & 1.3147 \\
270/365 & 3.8720 & 3.7975 & 3.8632 & 3.8682 & 3.8850 \\
\bottomrule
\end{tabular}


\begin{tabular}{lrrrrr} 
\multicolumn{6}{l}{Table 4}  \\
\toprule
 Stock &  \multicolumn{5}{l}{American calls}  \\
   \multicolumn{6}{l}{$\tau=0.5$ year,  $r=0.05$, $d=0.03$, $\sigma^2=0.16$, $\lambda=1$, $n=150$, $K =100$ } \\
\cmidrule(lr){1-6}
 &  \multicolumn{1}{l}{Simonato} &  \multicolumn{1}{l}{Dai} &  \multicolumn{1}{l}{HS} &  \multicolumn{1}{l}{HScut}  &  \multicolumn{1}{l}{Benchmark} \\
\midrule
\multicolumn{6}{l}{Panel A: $\gamma'=0.0000$, $\delta=0.1980 $, } \\
80 & 4.0966 & 4.0839 & 4.0956 & 5.0956  & 4.0500 \\
100 & 12.7026 & 12.6936 &  12.6912& 12.6912  & 12.6800 \\
120  & 26.2072 & 26.2035 & 26.2015 & 26.2015  & 26.2200 \\
\cmidrule(lr){2-6}
\multicolumn{6}{l}{Panel B: $\gamma'=0.0488$, $\delta=0.1888 $, } \\
80 & 4.2107 & 4.1867 &4.1983 & 4.1983  & 4.1200 \\
100 & 12.7409 & 12.7344 & 12.7312 & 12.7312 & 12.6800 \\
120 & 26.1668 & 26.1624 & 26.1591 & 26.1591 & 26.1400 \\
\cmidrule(lr){2-6}
\multicolumn{6}{l}{Panel C: $\gamma'=-0.0513$, $\delta=0.2082 $, } \\
80 &  4.0685 & 4.0722 & 4.0836 & 4.0836  & 4.0700 \\
100 &  12.8002  & 12.7887  & 12.7868 & 12.7868  & 12.8300 \\
120 &  26.3915 & 26.3809 & 26.3794 & 26.3794  & 26.4600 \\
\bottomrule
\end{tabular}

\normalsize

\section{Conclusion}

In this paper we investigate the option evaluation problem when the underlying price follows the bivariate discrete version of a jump diffusion process \textit{à la} Merton. We provide the estimation error due to the cutting of the bivariate lattice which describes the evolution of the underlying price.
This implies that we can make this error arbitrarily small while keeping proportional to $\ln n$ the number of branches sprouting from the discretisation of the jump part. We provide explicit formulas for finding the appropriate cutting, given a desired error.
Numerical tests show that even stronger pruning is feasible, with no appreciable variations in the error.
 Our method allows to reach results as precise as those of the HS procedure, while reducing the computational complexity to $O(n \ln n)$ in the European case and to $O(n^2 \ln n)$ in the American case when working on a bivariate tree.

\section*{References}

\bibliography{mybibfile}

\end{document}